\input ./style/arxiv-vmsta.cfg
\documentclass[numbers,compress]{vmsta}

\volume{2}
\pubyear{2015}
\firstpage{51}
\lastpage{65}
\doi{10.15559/15-VMSTA22}

\startlocaldefs

\allowdisplaybreaks
\endlocaldefs

\begin{document}
\begin{frontmatter}

\title{Autoregressive approaches to import--export time series I: basic
techniques}
\author[a]{\inits{L.}\fnm{Luca}\snm{Di Persio}}\email{dipersioluca@gmail.com}


\address[a]{Dept. Informatics, University of Verona,
strada le Grazie 15, 37134, Italy}

\markboth{L. Di Persio}{Autoregressive approaches to import--export
time series I: basic techniques}

\begin{abstract}
This work is the first part of a project dealing with an in-depth study
of effective techniques used in econometrics in order to make accurate
forecasts in the concrete framework of one of the major economies of
the most productive Italian area, namely the province of Verona.
In particular, we develop an approach mainly based on vector
autoregressions, where lagged values of two or more variables are
considered, Granger causality, and the stochastic trend approach useful
to work with the cointegration phenomenon.
Latter techniques constitute the core of the present paper, whereas in
the second part of the project, we present how these approaches can be
applied to economic data at our disposal in order to obtain concrete
analysis of import--export behavior for the considered productive area
of Verona.
\end{abstract}

%
\received{9 February 2015}
%
\revised{8 April 2015}
\accepted{8 April 2015}
\publishedonline{20 April 2015}
\end{frontmatter}

\begin{section}{Introduction}
The analysis of time series data constitutes a key ingredient in
econometric studies.
Last years have been characterized by an increasing interest toward the
study of econometric time series. Although various types of regression
analysis and related forecast methods are rather old, the worldwide
financial crisis experienced by markets starting from last months of
2007, and which is not yet finished, has put more attention on the subject.
Moreover, analysis and forecast problems have become of great momentum
even for medium and small enterprizes since their economic
sustainability is strictly related to the propensity of a bank to give
credits at reasonable conditions.

In particular, great efforts have been made to read economic data not
as monads, but rather as constituting pieces of a whole. Namely, new
techniques have been developed to study interconnections and
dependencies between different factors characterizing the economic
history of a certain market, a given firm, a specified industrial area,
and so on.
From this point of view, methods such as the vector autoregression, the
cointegration approach, and the copula techniques have been benefitted
by new research impulses.

A challenging problem is then to apply such instruments in concrete
situations and the problem becomes even harder if we take into account
the economies are hardly hit by the aforementioned crisis.
A particularly important case study is constituted by a close analysis
of import--export time series. In fact, such an information, spanning
from countries to small firms, has the characteristic to provide highly
interesting hints for people, for example, politicians or CEOs, to
depict future economic scenarios and related investment plans for the
markets in which they are involved.

Exploiting precious economic data that the Commerce Chamber of
Verona\break
Province has put at our disposal, we successfully applied some of the
relevant approaches already cited to find dependencies between economic
factor characterizing the Province economy and then to make effective
forecasts, very close to the real behavior of studied markets.

For completeness, we have split our project into two parts, namely the
present one, which aims at giving a self-contained introduction to the
statistical techniques of interest, and the second one, where the
Verona import--export case study have been treated in detail.

In what follows, we first recall univariate time series models, paying
particular attention to the AR model, which relates a time series to
its past values. We will explain how to make predictions, by using
these models, how to choose the delays, for example, using the Akaike
and Bayesian information crtiteria (AIC, resp. BIC), and how to behave
in the presence of trends or structural breaks.
Then we move to the vector autoregression (VAR) model, in which lagged
values of two or more variables are used to forecast future values of
these variables. Moreover, we present the Granger causality, and, in
the last part, we return to the topic of stochastic trend introducing
the phenomenon of cointegration. 
\end{section}

\section{Univariate time-series models}

Univariate models have been widely used for short-run forecast (see,
e.g., \cite[Examples of Chapter 2]{rif1}. In what follows, we recall
some of these techniques, focusing ourselves particularly on the
analysis of autoregressive (AR) processes, moving average (MA)
processes, and a combination of both types, the so-called ARMA
processes; for further details, see, for example, \cite{rif6, rif7,
rif8} and references therein.

The observation on the time-series variable $Y$ made at date $t$ is
denoted by~$Y_t$, whereas $T \in\mathbb{N}^+$ indicates the total
number of observations. Moreover, we denote the $ j$th lag of a time
series $ \{ Y_t  \}_{t=0, \ldots, T}$ by $Y_{t-j}$ (the value of
the variable $Y$ $j$ periods ago); similarly, $Y_{t+j}$ denotes the
value of $Y$ $j$ periods to the future, where, for any fixed $ t \in
\{ 0, \ldots, T \}, j $ is such that $ j \in\mathbb{N}^+ ,\ t-j \geq
0, $ and $ t+j \leq T$. The $ j$th autocovariance of a series $Y_t$ is
the covariance between $Y_t$ and its $j$th lag, that~is,
$\mathit{autocovariance}_j = \sigma_j : = \operatorname{cov}(Y_t,Y_{t-j})$,
whereas the $ j$th autocorrelation coefficient is the correlation
between $Y_t$ and $Y_{t-j}$, thats is,
$\mathit{autocorrelation}_j = \rho_j : = \operatorname{corr}(Y_t,Y_{t-j}) = \frac{
\operatorname{cov}(Y_t,Y_{t-j}) }{\sqrt{\operatorname{var}(Y_t)\operatorname{var}(Y_{t-j})} }$.
When the~average and variance of a variable are unknown, we can
estimate them by taking a random sample of $n$ observations.
In a simple random sample, $n$ objects are drawn at random from a
population, and each object is equally likely to be drawn. The value of
the random variable $Y$ for the $i$th randomly drawn object is denoted
$Y_i$. Because each object is equally likely to be drawn and the
distribution of $Y_i$ is the same for all $i$ , the random variables
$Y_1, \ldots, Y_n$ are independent and identically distributed
(i.i.d.). Given a variable $Y$, we denote by $\overline Y$ its sample
average with respect to the $n$ observations $Y_1, \ldots, Y_n$, thats is,
$
\overline Y = \frac{1}{n} ( Y_1 + Y_2 + \cdots + Y_n) = \frac{1}{n}
\sum_{i=1}^n Y_i
$, whereas we define the related sample variance by
$
s_Y^2 : = \frac{1}{n-1} \sum_{i=1}^n (Y_i - \overline Y)^2.
$ The $ j$th autocovariances, resp. autocorrelations, can be estimated
by the $ j$th sample autocovariances, resp. autocorrelations, as follows:
$
\widehat{\sigma_j} : = \frac{1}{T} \sum_{t=j+1}^T (Y_t-\overline
Y_{j+1,T})(Y_{t-j}-\overline Y_{1,T-j})$, resp.
$
\widehat{\rho_j} : = \frac{\widehat{\sigma_j}}{ s_Y^2}
$, where $\overline Y_{j+1,T}$ denotes the sample average of $ Y_t$
computed over the observations $ t = j+1,\dots,T$.
Concerning forecast based on regression models that relates a time
series variable to its past values, for completeness, we shall start
with the first-order autoregressive process, namely the AR(1) model,
which uses $Y_{t-1}$ to forecast $Y_t$.
A systematic way to forecast is to estimate an ordinary least squares
(OLS) regression. The OLS estimator chooses the regression coefficients
so that the estimated regression line is as close as possible to the
observed data, where the closeness is measured by the sum of the
squared mistakes made in predicting $Y_t$ given $Y_{t-1}$.
Hence, the AR(1) model for the series $Y_t$ is given by
\begin{equation}
\label{AR(1)} Y_t = \beta_0 + \beta_1
Y_{t-1} + u_t ,
\end{equation}
where $\beta_0$ and $\beta_1$ are the regression coefficients. In this
case, the intercept $\beta_0$ is the value of the regression line when
$Y_{t-1} = 0$, the slope $\beta_1$ represents the change in $Y_t$
associated with a unit change in $Y_{t-1}$, and $u_t$ denotes the error
term whose nature will be later clarified.
Let us assume that the value $Y_{t_0}$ of the time series $Y_t$ at
initial time $t_0$ is given; then
$
Y_{t_0+1} = \beta_0 + \beta_1 Y_{t_0} + u_{t_0+1}$, so that iterating
relation \eqref{AR(1)} up to order $\tau> 0$ , we get
\begin{displaymath}
\begin{split} Y_{t_0+\tau} &= \bigl( 1 + \beta_1 +
\beta_1^2 + \cdots+ \beta_1^{\tau-1}
\bigr) \beta_0 + \beta_1^\tau Y_{t_0}\\
& \quad + \beta_1^{\tau-1} u_{t_0+1} +
\beta_1^{\tau-2} u_{t_0+2} + \cdots+
\beta_1 u_{t_0+\tau- 1} + u_{t_0+\tau}\\
&= \beta_1^\tau Y_{t_0} + \frac{1 - \beta_1^\tau}{ 1 - \beta_1}
\beta_0 + \sum_{j=0}^{\tau-1}
\beta_1^j u_{t_0 + \tau- j}.
\end{split} %
\end{displaymath}
Hence, taking $t = t_0 + \tau$ with $t_0 = 0$, we obtain
\begin{equation}
\label{AR(1)2} Y_t = \beta_1^t
Y_0 + \frac{1 - \beta_1^t}{ 1 - \beta_1} \beta_0 + \sum
_{j=0}^{t-1} \beta_1^j
u_{t - j} .
\end{equation}
A time series $Y_t$ is called \emph{stationary} if its probability
distribution does not change over time,
that is, if the joint distribution of $(Y_{s+1}, Y_{s+2}, \dots, Y_{s+T})$
does not depend on $s$; otherwise, $Y_t$ is said to be \emph{nonstationary}.
In \eqref{AR(1)2}, the process $Y_t$ consists of both time-dependent
deterministic and stochastic parts, and, thus, it cannot be
stationary.

Formally, the process with stochastic initial conditions results from
\eqref{AR(1)2} if and only if $|\beta_1| < 1$. It follows that if $\lim_{t_0 \to- \infty} Y_{t_0}$ is bounded, then, as $ t_0 \rightarrow-
\infty$, we have
\begin{equation}
\label{AR(1)3} Y_t = \frac{\beta_0}{ 1 - \beta_1} + \sum
_{j=0}^{\infty} \beta_1^j
u_{t
- j};
\end{equation}
see, for example, \cite[Chap. 2.1.1]{rif1}. Equation \eqref{AR(1)3}
can be rewritten by means of the lag operator, which acts as follows:
$
L Y_t = Y_{t-1},\, L^2 Y_t = Y_{t-2}, \ldots, L^k Y_t = Y_{t-k} $,
so that Eq.~\eqref{AR(1)} becomes $
(1-\beta_1 L) Y_t = \beta_0 + u_t $. Assuming that $E[u_t] = 0$ for all
$t$, we have
\begin{align*}
E[Y_t] &= E \Biggl[ \frac{\beta_0}{ 1 - \beta_1} + \sum
_{j=0}^{\infty} \beta_1^j
u_{t - j} \Biggr] = \frac{\beta_0}{ 1 - \beta_1} + \sum
_{j=0}^{\infty} \beta_1^j E[
u_{t - j}] =\frac{\beta_0}{ 1 - \beta_1} = \mu,\\
 V[Y_t] &= E \biggl[ \biggl( Y_t
- \frac{\beta_0}{ 1 - \beta_1} \biggr)^2 \biggr] = E \Biggl[ \Biggl( \sum
_{j=0}^{\infty} \beta_1^j
u_{t - j} \Biggr)^2 \Biggr]\\
& = E\bigl[\bigl(u_t+\beta_1 u_{t-1}+
\beta_1^2 u_{t-2} + \cdots\bigr)^2
\bigr]\\
&= E\bigl[u_t^2+\beta_1^2
u_{t-1}^2+\beta_1^4
u_{t-2}^2 + \cdots+ 2 \beta_1 u_t
u_{t-1} + 2 \beta_1^2 u_t
u_{t-2} +\cdots\bigr]\\
&=\sigma^2 \bigl(1+ \beta_1^2+
\beta_1^4+ \cdots\bigr)=\frac{\sigma^2}{1-\beta
_1^2} ,
\end{align*}
where we have used that $E[u_t u_s] = 0$ for $t \neq s$ and $|\beta_1|
< 1$. Hence, both the mean and variance are constants, and thus the
covariances are given by
\begin{displaymath}
\begin{split} \operatorname{Cov}[Y_t, Y_{t-1}] &= E \biggl[
\biggl( Y_t - \frac{\beta_0}{ 1 - \beta
_1} \biggr) \biggl( Y_{t-1}-
\frac{\beta_0}{ 1 - \beta_1} \biggr) \biggr]\\
&= E\big[\bigl(u_t+\beta_1 u_{t-1}+\cdots+
\beta_1^\tau u_{t-\tau} + \cdots\bigr)\\
& \quad\xch{\times}{} \bigl(u_{t-\tau}+\beta_1 u_{t-\tau-1}+
\beta_1^2 u_{t-\tau-2} + \cdots\bigr)\big]\\
&= E\bigl[\bigl(u_t+\beta_1 u_{t-1}+\cdots+
\beta_1^{\tau-1} u_{t-\tau-1}\\
& \quad +\beta_1^\tau\bigl(u_{t-\tau}+
\beta_1 u_{t-\tau-1}+\beta_1^2
u_{t-\tau-2} + \cdots\bigr)\bigr)\\
& \quad\xch{\times}{} \bigl(u_{t-\tau}+\beta_1 u_{t-\tau-1}+
\beta_1^2 u_{t-\tau-2} + \cdots\bigr)\bigr]\\
&=\beta_1^\tau E\bigl[\bigl(u_{t-\tau}+
\beta_1 u_{t-\tau-1}+\beta_1^2 u_{t-\tau
-2} + \cdots\bigr)^2\bigr]=\beta_1^\tau
V[Y_{t-\tau}]\\
&=\beta_1^\tau\frac{\sigma^2}{1-\beta_1^2} =: \gamma(\tau) .
\end{split} %
\end{displaymath}
The previous AR(1) can be generalized by considering arbitrary but
finite order \mbox{$p>1$}. In particular ,
an AR($p$) process can be described by the equation
\begin{equation}
\label{AR(p)} Y_t = \beta_0 + \beta_1
Y_{t-1} + \beta_2 Y_{t-2} + \cdots+
\beta_p Y_{t-p} + u_t ,
\end{equation}
where $\beta_0, \ldots, \beta_p$ are constants, whereas $u_t$ is the
error term represented by a~random variable with zero mean and variance
$\sigma^2 > 0$. Using the lag operator, we can rewrite Eq.~\eqref
{AR(p)} as
$
(1-\beta_1 L- \beta_2 L^2 - \cdots- \beta_p L^p) Y_t = \beta_0 + u_t $.
In such a framework, it is standard to assume that the following four
properties hold (see, e.g., \cite[Chap.\ 14.4]{rif2}):
\begin{itemize}
\item$u_t$ has conditional mean zero, given all the regressors, that
is,\break
$E(u_t|Y_{t-1}, Y_{t-2}, \ldots) = 0$, which implies that the best
forecast of $Y_t$ is given by the $\mathrm{AR}(p)$ regression.
\item$Y_i$ has a stationary distribution, and $Y_i$, $Y_{i-j}$ are
assumed to become independent as $j$ gets large. If the time-series
variables are nonstationary, then the forecast can be biased and
inefficient, or conventional OLS-based statistical inferences can be misleading.
\item All the variables have nonzero finite fourth moments.
\item There is no perfect multicollinearity, namely it is not true
that, given a~certain regressor, it is a perfect linear function of the
variables.
\end{itemize}

\subsection{Forecasts}
In this section, we show how the previously introduced class of models
can be
used to predict the future behavior of a certain quantity of interest.
If $Y_t$ follows the $\mathrm{AR}(p)$ model and $\beta_0 , \beta_1 , \ldots,
\beta_p$ are unknown, then the forecast of $Y_{T+1}$ is given by $\beta
_0+\beta_1 Y_T+\beta_2 Y_{T-1} +\cdots+ \beta_p Y_{T-p+1}$. Forecasts
must be based on estimates of the coefficients $\beta_i$ by using the
OLS estimators based on historical data. Let $\hat Y_{T+1}$ denote the
forecast of $Y_{T+1}$ based on $Y_T,Y_{T-1},\ldots$:
\begin{displaymath}
\hat{Y}_{T+1|T}= \hat{\beta}_0 + \hat{\beta}_1
Y_T + \hat{\beta}_2 Y_{T-1}+ \cdots+\hat{
\beta}_p Y_{T-p+1} .
\end{displaymath}
Then such a forecast refers to some data beyond the data set used to
estimate the regression, so that the data on the actual value of the
forecasted dependent variable are not in the sample used to estimate
the regression. Forecasts and forecast error pertain to
``out-of-sample'' observations.

The forecast error is the mistake made by the forecast; this is the
difference between the value of $Y_{T+1}$ that actually occurred and
its forecasted value
$
\mbox{forecast}\break \mbox{error} : = Y_{T+1} - \hat{Y}_{T+1|T} $.

The root mean squared forecast error RMSFE is a measure of the size of
the forecast error
$\mathit{RMSFE} =\sqrt{E[ (Y_{T+1} - \hat{Y}_{T+1|T})^2]}$,
and it is characterized by two sources of error: the error arising
because future values of $ u_t$ are unknown and the error in estimating
the coefficients $\beta_i$. If the first source of error is much larger
than the second, the RMSFE is approximately $\sqrt{\mathrm{var}(u_t)}$, the
standard deviation of the error $u_t$, which is estimated by the
standard error of regression (SER).
One useful application used in time-series forecasting is to test
whether the lags of one regressor have useful predictive content. The
claim that a variable has no predictive content corresponds to the null
hypothesis that the coefficients on all lags of that variable are zero.
Such a hypothesis can be checked by the so-called Granger causality
test (GCT), a type of F-statistic approach used to test joint
hypothesis about regression coefficients. In particular, the GCT method
tests the hypothesis that the coefficients of all the values of the
variable in
$
Y_t = \beta_0 + \beta_1 Y_{t-1} + \beta_2 Y_{t-2} + \cdots+ \beta_p
Y_{t-p} + u_t $,
namely the coefficients of $ Y_{t-1} , Y_{t-2} , \ldots, Y_{t-p} $,
are zero, and hence this null hypothesis implies that such regressors
have no predictive content for $Y_t$.
\subsection{Lag length selection} \label{lag1}
Let us recall relevant statistical methods used to optimally choose the
number of lags in an autoregression model; in particular, we focus our
attention on the \emph{Bayes} method (BIC) and on the \emph{Akaike}
method (AIC); for more details, see, for example, \cite[Chap.~14.5]{rif2}.
The BIC method is specified by
\begin{equation}
\label{BIC} \mathrm{BIC}(p)=\ln \biggl( \frac{\mathrm{SSR}(p)}{T} \biggr) + (p+1)
\frac{\ln T}{T} ,
\end{equation}
where $\mathrm{SSR}(p)$ is the \emph{sum of squared residuals} of the estimated
$\mathrm{AR}(p)$. The $\mathit{BIC}$ estimator of $p$ is the value that minimizes
$\mathrm{BIC}(p)$ among all the possible choices. In the first term of Eq.~\eqref
{BIC}, the sum of squared residuals necessarily decreases when adding a
lag. In contrast, the second term is the number of estimated regression
coefficients times the factor $(\ln T)/T$, so this term increases when
adding a lag. This implies that the $\mathit{BIC}$ trades off these two aspects.
The AIC approach is defined by
\begin{displaymath}
\mathrm{AIC}(p)=\ln \biggl( \frac{\mathrm{SSR}(p)}{T} \biggr) + (p+1)\frac{2}{T} ,
\end{displaymath}
and hence the main difference between the $\mathit{AIC}$ and $\mathit{BIC}$ is that the
term $\ln(T)$ in the $\mathit{BIC}$ is replaced by $2$ in the $\mathit{AIC}$, so the
second term in the $\mathit{AIC}$ is smaller. But the second term in the $\mathit{AIC}$
is not large enough to assure choosing the correct length, so this
estimator of $p$ is not consistent. We recall that an estimator is
consistent if, as the size of the sample increases, its probability
distribution concentrates at the value of the parameter to be
estimated. So, the BIC estimator $\hat{p}$ of the lag length in an
autoregression is correct in large samples, that is, $\operatorname{Pr}(\hat{p} = p)
\to1$. This is not true for the AlC estimator, which can overestimate
$p$ even in large samples; for the proof, see, for example, \cite
[Appendix 14.5]{rif2}.
\subsection{Trends} \label{trends}
A further relevant topic in econometric analysis is constituted by
nonstationarities that are due to trends and breaks. A trend is a
persistent long-term movement of a variable over time. A time-series
variable fluctuates around its trend. There are two types of trends,
deterministic and stochastic. A \emph{deterministic trend} is a
nonrandom function of time. In contrast, a stochastic trend is
characterized by a random behavior over time. Our treatment of trends
in economic time series focuses on stochastic trend. One of the
simplest models of time series with stochastic trend is the
one-dimensional \emph{random walk} defined by the relation
$
Y_t = Y_{t-1} + u_t
$, where $u_t$ is the error term represented by a normally distributed
random variable with zero mean and variance $\sigma^2 > 0$. In this
case, the best forecast of tomorrow's value is its value today. A
extension of the latter is the \emph{random walk with drift} defined by
$
Y_t = \beta_0 + Y_{t-1} + u_t ,\  \beta_0 \in\mathbb{R}$,
where the best forecast is the value of the series today plus the drift
$\beta_0$.
A~random walk is nonstationary because the variance of a random walk
increases over time, so the distribution of $Y_t$ changes over time.
In fact, since $u_t$ is uncorrelated with $Y_{t-1}$, we have $\mathrm{var}(Y_t)
= \mathrm{var}(Y_{t-1}) + \mathrm{var}(u_t)$ with $\mathrm{var}(Y_t) = \mathrm{var}(Y_{t-1})$ if and only
if $\mathrm{var}(u_t)=0$. The random walk is a particular case of an $\mathrm{AR}(1)$
model with $\beta_1 = 1$. If $|\beta_1|<1$ and $u_t$ is stationary,
then $Y_t$ is stationary. The condition for the stationarity of an
$\mathrm{AR}(p)$ model is that the roots of
$1-\beta_1 z - \beta_2 z^2 - \beta_3 z^3 - \cdots- \beta_p z^p =0 $
are greater than one in absolute value. If an $\mathrm{AR}(p)$ has a root equal
to one, then we say that the series has a \emph{unit root} and a \emph
{stochastic trend}. Stochastic trends usually bring many issues, for
example, the autoregressive coefficients are biased toward zero.
Because $Y_t$ is nonstationary, the assumptions for time-series
regression do not hold, and we cannot rely on estimators and test
statistics having their usual large-sample normal distributions; see,
for example, \cite[Chap.\  3.2]{rif2}.
In fact, the OLS estimator of the autoregressive coefficient $\hat{\beta
}_1$ is consistent, but it has a nonnormal distribution; then the
asymptotic distribution of $\hat{\beta}_1$ is shifted toward zero.
Another problem caused by stochastic trend is the nonnormal
distribution of the t-statistic, which means that conventional
confidence intervals are not valid and hypothesis tests cannot be
conducted as usual. The t-statistic is an important example of a test
statistic, namely of a statistic used to perform a~hypothesis test.
A statistical hypothesis test can make two types of mistakes: a \emph
{type I error}, in which
the null hypothesis is rejected when, in fact, it is true, and
a \emph{type II error}, in which
the null hypothesis is not rejected when, in fact, it is false. The
prespecified rejection
probability of a statistical hypothesis test when the null hypothesis
is true, that
is, the prespecified probability of a type I error, is called the \emph
{significance level} of
the test. The \emph{critical value} of the test statistic is the value
of the statistic for which
the test just rejects the null hypothesis at the given significance level.
The \emph{p-value} is the probability of obtaining a test statistic, by
random sampling
variation, at least as adverse to the null hypothesis value as is the
statistic actually
observed, assuming that the null hypothesis is correct. Equivalently,
the $p$-value is
the smallest significance level at which you can reject the null hypothesis.
The value of the t-statistic is
\begin{displaymath}
t = \frac{\mathit{estimator} - \mathit{hypothesized} \ \mathit{value}}{\mathit{standard}
\ \mathit{error} \ \mathit{of} \ \mathit{the}
\ \mathit{estimator}}
\end{displaymath}
and is well approximated by the standard normal distribution when $n$
is large because of the central limit theorem (see, e.g., \cite[Chap.\
4.3]{rif4}). Moreover, stochastic trends can lead two time series to
appear related when they are not, a problem called \emph{spurious
regression} (see, e.g., \cite[Chap.\ 2]{rif3} for examples). For the
$\mathrm{AR}(1)$ model, the most commonly used test to determine stochastic
trends, is the Dickey--Fuller test (see, e.g., \cite[Chap.\ 3]{rif3}
for details. For this test, we first subtract $Y_{t-1}$ from both sides
of the equation $Y_t = \beta_0 + \beta_1 Y_{t-1} + u_t$. Then we assume
that the following hypothesis test holds:
\begin{displaymath}
H_0 : \delta= 0 \quad \mathrm{versus} \quad H_1 : \delta< 0
\qquad \mathrm{in} \ Y_t - Y_{t-1} = \Delta Y_t =
\beta_0 + \delta Y_{t-1} + u_t
\end{displaymath}
with $\delta= \beta_1-1$. For an $AR(p)$ model, it is standard to use
the augmented Dickey--Fuller test (ADF), which tests the null
hypothesis $H_0 : \delta= 0$ against the one-side alternative $H_1 :
\delta< 0$ in the regression
\begin{displaymath}
\Delta Y_t = \beta_0 + \delta Y_{t-1} +
\gamma_1 \Delta Y_{t-1} + \gamma _2 \Delta
Y_{t-2}+ \cdots+ \gamma_p \Delta Y_{t-p} +
u_t
\end{displaymath}
under the null hypothesis. Let us note that since $Y_t$ has a
stochastic trend, it follows that, under the alternative hypothesis,
$Y_t$ is stationary. The ADF statistic is the OLS t-statistic testing
$\delta=0$. If, instead, the alternative hypothesis is that $Y_t$
is
stationary around a deterministic linear time trend, then this trend
$t$ must be added as an additional regressor. In this case, the
Dickey--Fuller regression becomes\vadjust{\eject}
\begin{displaymath}
\Delta Y_t = \beta_0 + \alpha t + \delta
Y_{t-1} + \gamma_1 \Delta Y_{t-1} +
\gamma_2 \Delta Y_{t-2}+ \cdots+ \gamma_p \Delta
Y_{t-p} + u_t ,
\end{displaymath}
and we test for $\delta= 0$. The ADF statistic does not have a normal
distribution, and hence different critical values have to be used.

\subsection{Breaks} \label{breaks}
A second type of nonstationarity arises when the regression function
changes over the course of the sample. In economics, this can occur for
a variety of reasons, such as changes in economic policy, changes in
the structure of the economy, or an invention that changes a specific
industry. These breaks cannot be neglected by the regression model. A
problem caused by breaks is that the OLS regression estimates over the
full sample will estimate a relationship that holds ``on average,'' in
the sense that the estimate combines two different periods, and this
leads to poor forecast. There are two types of testing for breaks:
testing for a break at a known date and for a break at an unknown break
date. We consider the first option for an $\mathrm{AR}(p)$ model. Let $\tau$
denote the hypothesized break date, and let $D_t (\tau)$ be the binary
variable such that $D_t (\tau) = 0$ if $t>\tau$ and $D_t (\tau) = 1$ if
$t<\tau$. Then the regression including the binary break indicator and
all interaction terms reads as follows:
\begin{displaymath}
\begin{split} Y_t &= \beta_0 +
\beta_1 Y_{t-1} + \beta_2 Y_{t-2} +
\cdots+ \beta_p Y_{t-p}+ \gamma_0
D_t (\tau)
\\
& \quad {} + \gamma_1 \bigl[D_t (\tau) \times
Y_{t-1} \bigr] + \gamma_2 \bigl[D_t (\tau )
\times Y_{t-2} \bigr] + \cdots+ \gamma_p
\bigl[D_t (\tau) \times Y_{t-p} \bigr] + u_t
\end{split} %
\end{displaymath}
under the null hypothesis of no breaks, $\gamma_0 = \gamma_1 = \gamma_2
= \cdots= \gamma_p = 0 $. Under the alternative hypothesis that there
is a break, the regression function is different before and after the
break date $\tau$, and we can use the F-statistic performing the
so-called the Chow test (see, e.g., \cite[Chap.\  5.3.3]{rif1}). If we
suspect a break between two dates $\tau_0$ and~$\tau_1$, the Chow test
can be modified to test for breaks at all possible dates $\tau$ between
$\tau_0$ and $\tau_1$, then using the largest of the resulting
F-statistics to test for a break at an unknown date. The latter
technique is called the {\it Quandt likelihood ratio statistic} (QLR)
(see, e.g., \cite[Chap.\ 14.7]{rif2}).
Because the QLR statistic is the largest of many F-statistics, its
distribution is not the same as that of an individual F-statistic;
also, the critical values for the QLR statistic must be obtained from a
special distribution.
\section{MA and ARMA}
In the following, we consider finite-order
moving-average (MA) processes (see, e.g., \cite[Chap.\  2.2]{rif1}).
The \emph{moving-average process of order $q$}, MA($q$), is defined by
$
Y_t = \alpha_0 +u_t - \alpha_1 u_{t-1} - \alpha_2 u_{t-2} - \cdots-
\alpha_q u_{t-q}$;
equivalently, by using the lag operator we get
$
Y_t - \alpha_0 = (1-\alpha_1 L- \alpha_2 L^2 - \cdots- \alpha_q L^q) u_t$.
Every finite MA($q$) process is stationary, and we have
\begin{itemize}
\item$ E[Y_t] = \alpha_0, $
\item$ V[Y_t] = E[(Y_t - \alpha_0)^2] = (1+ \alpha_1^2 + \alpha_2^2 +
\cdots+ \alpha_q^2 )\sigma^2, $
\item$ \operatorname{Cov}[Y_t, Y_{t+\tau}] \ = E[(Y_t - \alpha_0)(Y_{t+\tau} - \alpha
_0)] $ \\
\hspace*{2.16cm}$ = E[u_t (u_{t+\tau} - \alpha_1 u_{t+\tau-1} - \cdots-
\alpha_q u_{t+\tau-q}) $ \\
\hspace*{2.5cm}$ {}- \alpha_1 u_{t-1} (u_{t+\tau} - \alpha_1 u_{t+\tau
-1} - \cdots- \alpha_q u_{t+\tau-q})$ \\
\hspace*{2.5cm}$ \dots- \alpha_q u_{t-q} (u_{t+\tau} - \alpha_1
u_{t+\tau-1} - \cdots- \alpha_q u_{t+\tau-q})]. $
\end{itemize}
Combining both an autoregressive (AR) term of order $p$ and a
moving-average (MA) term of order $q$, we can define the process
denoted as ARMA($p,q$) and represented by
\begin{displaymath}
Y_t = \beta_0 +\beta_1 Y_{t-1} +
\cdots+ \beta_p Y_{t-p} + u_t - \alpha
_1 u_{t-1} - \cdots- \alpha_q u_{t-q}
;
\end{displaymath}
again, exploiting the lag operator, we can write
\begin{displaymath}
\begin{split} \bigl(1-\beta_1 L- \beta_2
L^2 - \cdots- \beta_p L^p\bigr)
Y_t &= \beta_0 + \bigl(1-\alpha_1 L- \alpha_2
L^2 - \cdots- \alpha_q L^q\bigr)
u_t ,\\
\beta(L) Y_t &= \beta_0 + \alpha(L) u_t .
\end{split} %
\end{displaymath}

\section{Vector autoregression}
In what follows, we focus our study on the so-called vector
autoregression (VAR) econometric model, also using some remarks on the
relation between the univariate time series models described in the
first part, and the set of simultaneous equations systems of
traditional econometrics characterizing the VAR approach (see, e.g.,
\cite[Chap.\ 2]{rif5}).
\subsection{Representation of the system}
We have so far considered forecasting a single variable. However, it is
often necessary to allow for a multidimensional statistical analysis if
we want to forecast more than one-parameter dynamics. This section
introduces a model for forecasting multiple variables, namely the
vector autoregression (VAR) model, in which lagged values of two or
more variables are used to forecast their future values. We start with
the autoregressive representation in a VAR model of order $p$, denoted
by VAR($p$), where each component depends on its own lagged values up
to $p$ periods and on the lagged values of all other variables up to
order $p$. It follows that the main idea behind the VAR model is to
know how new information, appearing at a certain time point and
concerning one of the observed variables, is processed in the system
and which impact it has over time not only for this particular variable
but also for the other system parameters. Hence, a VAR($p$) model is a
set of $k$ time-series regressions $(k \in\mathbb{N}^+$) in which the
regressors are lagged values of all $k$ series and the number of lags
equals~$p$ for each equation.
In the case of two time series variables, say, $Y_t$ and $X_t$, the
VAR($p$) consists of two equations of the form
\begin{equation}
\label{VAR} %
\begin{cases}
Y_t = \beta_{10} + \beta_{11} Y_{t-1} + \cdots+ \beta_{1p} Y_{t-p} +
\gamma_{11} X_{t-1} + \cdots+ \gamma_{1p} X_{t-p} + u_{1t},\\X_t = \beta
_{20} + \beta_{21} Y_{t-1} + \cdots+ \beta_{2p} Y_{t-p} + \gamma_{21}
X_{t-1} + \cdots+ \gamma_{2p} X_{t-p} + u_{2t},
\end{cases} %
\end{equation}
where the $\beta$s and the $\gamma$s are unknown coefficients, and
$u_{1t}$ and $u_{2t}$ are error terms represented by normally
distributed random variables with zero mean and variance $\sigma_i^2 >
0$. The VAR assumptions are the same as those for the time-series
regression defining AR models and applied to each equation; moreover,
the coefficients of each VAR are estimated by means of the OLS
approach. The reduced form of a vector autoregression of order$p$ is
defined as \mbox{$Z_t = \delta+ A_1 Z_{t-1} + A_2 Z_{t-2} + \cdots+ A_p Z_{t-p} + U_t$},
where $A_i,\, i = 1, \ldots, p$, are \hbox{$k$-dimensional}
quadratic matrices, $U$ represents the $k$-dimensional vector of
residuals at time~$t$, and \(\delta\) is the vector of constant terms.
System~\eqref{VAR} can be rewritten compactly as
$A_p(L) Z_t = \delta+ U_t$, where\vadjust{\eject}
$ A_p(L) =\break I_k - A_1L - A_2L^2 - \cdots- A_pL^p$, $E[U_t] = 0, \ E[U_t
U_t'] = \sigma_{uu}$, and $E[U_t U_s'] = 0$  for  $t \ne s$. Such
a~system is stable if and only if all included variables are
stationary, that is, if all roots of the characteristic equation of the
lag polynomial are outside the unit circle, namely $ \det( I_k - A_1z -
A_2z - \cdots- A_pz) \ne0$ for $|z| \leq1$ (for details, see, e.g.,
\cite[Chap.\ 4.1]{rif1}).
We use this condition because we saw in Section~\ref{trends} that
the condition for the stationarity of an $\mathrm{AR}(p)$ model is that the
roots of
$1-\beta_1 z - \beta_2 z^2 - \beta_3 z^3 - \cdots- \beta_p z^p =0 $
are greater than one in absolute value. If an $\mathrm{AR}(p)$ has a root equal
to one, we say that the series has a \emph{unit root} and a \emph
{stochastic trend}. Moreover, the previous system can be rewritten by
exploiting the MA representation as follows:
\begin{displaymath}
\begin{split} Z_t &= A^{-1}(L) \delta+
A^{-1}(L) U_t
\\
& = \mu+ U_t - B_1 U_{t-1} - B_2
U_{t-2} - B_3 U_{t-3} - \cdots
\\
& = \mu+ B(L) U_t \, \end{split} %
\end{displaymath}
with
\begin{displaymath}
\begin{split} B_0 = I_k \,, \quad& B(L) := I
- \sum_{j = 1}^{\infty} B_j
L^j \equiv A^{-1}(L) ,
\\
& \mu= A^{-1}(1) \delta= B(1) \delta. \end{split} %
\end{displaymath}
The autocovariance matrices are defined as
$\varGamma_Z(\tau) = E[(Z_t - \mu)(Z_{t-\tau} - \mu)']$;
without loss of generality, we set $\delta= 0$ and, therefore, $\mu=
0$, whence we obtain
\begin{displaymath}
\begin{split} E\bigl[Z_t Z_{t-\tau}'
\bigr] &= A_1 E\bigl[Z_{t-1} Z_{t-\tau}'
\bigr] + A_2 E\bigl[Z_{t-2} Z_{t-\tau}'
\bigr]
\\
&\quad +\cdots+ A_p E\bigl[Z_{t-p} Z_{t-\tau}'
\bigr] + E\bigl[U_t Z_{t-\tau}'\bigr] \quad
\end{split} %
\end{displaymath}
and, for $\tau\geq0$,
\begin{displaymath}
\begin{split}  \varGamma_Z(\tau) &= A_1
\varGamma_Z(\tau-1) + A_2 \varGamma_Z(
\tau-2) + \cdots + A_p \varGamma_Z(\tau-p) ,
\\
 \varGamma_Z(0) &= A_1 \varGamma_Z(-1) +
A_2 \varGamma_Z(-2) + \cdots+ A_p \varGamma
_Z(-p) + \varSigma_{uu}
\\
& = A_1 \varGamma_Z(1)'
+ A_2 \varGamma_Z(2)' + \cdots+
A_p \varGamma_Z(p)' +
\varSigma_{uu} . \end{split} %
\end{displaymath}

Since the autocovariance matrix entries link a variable with both its
delays and the remaining model variables,
we have that if 
the autocovariance between $X$ and $Y$ is positive, then $X$ tends to
move accordingly with $Y$ and vice~versa, whereas
if $X$ and $Y$ are independent, their autocovariance obviously equals zero.

\subsection{Determining lag lengths in VARs} \label{lag}
An appropriate method for the lag length selection of VAR is
fundamental to determine properties of VAR and related estimates.
There are two main approaches used for selecting or testing lag length
in VAR models. The first consists of rules of thumb based on the periodicity
of the data and past experience, and the second is based on formal
information criteria. VAR models typically include enough lags to capture
the full cycle of the data; for monthly data, this means that there is a
minimum of 12 lags, but we will also expect that there is some seasonality
that is carried over from year to year, so often lag
lengths of 13--15 months are used (see, e.g., \cite[Chap.\
2.5]{rif5}). For quarterly data, it is standard to use six lags.
This captures the cyclical components in the year and any residual seasonal
components in most cases. Usually, we decide to choose the number
of lags not exceeding $ kp + 1 < T$, where $k$ is the number of
endogenous variables, $p$ is the lag length, and $T$ is the total
number of
observations. We use this limitation because the estimate of all these
coefficients increases the amount of forecast estimation errors, which
can result in a~deterioration of the accuracy of the forecast itself.
The lag length in VAR can be formally determined using information
criteria; let $\hat{\varSigma}_{uu}$ be the estimate of the covariance
matrix with the $(i,j)$ element $\frac{1}{T} \sum_{t=1}^T \hat{u}_{it}
\hat{u}_{jt}$, where $\hat{u}_{it}$ is the OLS residual from the $j$th equation.
The BIC for the $k$th equation in a VAR model is
\begin{equation}
\label{BICVAR} \mathrm{BIC}(p) = \ln\bigl[\det(\hat{\varSigma}_{uu})\bigr] +
k(kp + 1) \frac{\ln T}{T},
\end{equation}
whereas the AIC is computed using Eq.~\eqref{BICVAR}, modified by
replacing the term $\ln T$ by~$2$. Among a set of candidate values of
$p$, the estimated lag length $\hat{p}$ is the value of $p$ that
minimizes BIC($p$).
\subsection{Multiperiod VAR forecast}
Iterated multivariate forecasts are computed using a VAR in much the
same way as univariate forecasts are computed using an autoregression.
The main new feature of a multivariate forecast is that the forecast of
one variable depends on the forecast of all variables in the VAR. To
compute multiperiod VAR forecasts $h$ periods ahead, it is necessary to
compute forecast of all variables for all intervening periods between
$T$ and $T + h$.
Then the following scheme applies: compute the one-period-ahead
forecast of all the variables in the VAR, then use those forecasts to
compute the two-period-ahead forecasts, and repeat the previous stops
until the desired forecast horizon. For example, the two-period-ahead
forecast of $Y_{T+2}$ based on the two-variable VAR($p$) in Eq.~\eqref
{VAR} is
\begin{align}
  \hat{Y}_{T+2 | T}& =  \hat{
\beta}_{10} + \hat{\beta}_{11} \hat{Y}_{T+1
| T} + \hat{
\beta}_{12} Y_T + \hat{\beta}_{13}
Y_{T-1} + \cdots+ + \hat {\beta}_{1p} Y_{T-p+2}\notag\\
&\quad+ \hat{\gamma}_{11} \hat{X}_{T+1 | T} + \hat{
\gamma}_{12} X_T + \hat {\gamma}_{13}
X_{T-1} + \cdots+ \hat{\gamma}_{1p} X_{T-p+2} ,\label{VARprev}
\end{align}
where the coefficients in \eqref{VARprev} are the OLS estimates of the
VAR coefficients.
\subsection{Granger causality}
An important question in multiple time series is to assign the value of
individual variables to explain the remaining ones in the considered
system of equations. An example is the value of a variable $Y_t$ for
predicting another variable $X_t$ in a dynamic system of equations or
understanding if the variable~$Y_t$ is informative about future values
of $X_t$. The answer is based on the determination of the so-called
Granger causality parameter for a time-series model (for details, see,
e.g., \cite[Chap.\ 2.5.4]{rif5}). To define the concept precisely,
consider the bivariate VAR
model for two variables $(Y_t , X_t)$ as in Eq.~ \eqref{VAR}. Using
this system of equations, Granger causality states that, for linear
models, $X_t$ Granger causes $Y_t$ if the behavior
of past $Y_t$ can better predict the behavior of~$X_t$ than the past
$X_t$ alone.
For the model in system \eqref{VAR}, if $X_t$ Granger\vadjust{\eject} causes $Y_t$,
then the coefficients for the past values of $X_t$ in the $Y_t$
equation are nonzero, that~is, $\gamma_{1i} \neq0$ for $i = 1, 2,
\ldots, p$. Similarly, if $Y_t$ Granger
causes $X_t$ in the $X_t$ equation, then the coefficients for the past
values of $Y_t$
are nonzero, that is, $\beta_{2i} \neq0$ for $i = 1, 2, \ldots, p$.
The formal testing for Granger causality is then done by using an
F test for the joint hypothesis that the possible causal variable
does not cause the other variable. We can specify the null hypothesis
for the
Granger causality test as follows.
\begin{displaymath}
\begin{split} & H_0{:}\ \boldsymbol{\mathrm{Granger \ noncausality}}
\ X_t  \mbox{ does  not  predict }  Y_t  \mbox{ if}
\\
&\ \ \ \ \ \ \ \ \ \ \ \ \ \ \ \ \ \ \ \ \ \ \gamma_{11} =
\gamma_{12} = \cdots= \gamma_{1p} = 0,
\\[6pt]
& H_1{:}\ \boldsymbol{\mathrm{Granger \ causality}} \ X_t
\mbox{ does   predict } Y_t \mbox{ if}
\\
&\ \ \ \ \ \ \ \ \ \ \ \ \ \ \ \ \ \ \ \ \ \ \gamma_{11} \neq0, \gamma
_{12} \neq0, \ldots, \mbox{ or }  \gamma_{1p} \neq0, \end{split}
\end{displaymath}
whereas the F test implementation is based on two models.
\begin{description}
 \item[Model 1] (unrestricted)\hfill

$ Y_t = \beta_{10} + \beta_{11} Y_{t-1}
+ \cdots+ \beta_{1p} Y_{t-p} + \gamma_{11}
X_{t-1} + \cdots+ \gamma_{1p} X_{t-p} +u_{1t} .$
\item[Model 2] (restricted)\hfill

$ Y_t = \beta_{10} + \beta_{11} Y_{t-1}
+ \cdots+ \beta_{1p} Y_{t-p} + u_{1t} .$
\end{description}

In the first model, we have $\gamma_{11} \neq0, \gamma_{12} \neq0,
\ldots, \gamma_{1p} \neq0$, so the variable $X_t$ compares in the
equation of $Y_t$, namely the values of $X_t$ are useful to
predict~$Y_t$. Instead, in the second model, $\gamma_{11} = \gamma_{12}
= \cdots= \gamma_{1p} = 0$, so $X_t$ does not Granger cause $Y_t$. The
test statistic has an $F$ distribution
with$(p , T - 2p -1)$ degrees of freedom:\vspace*{10pt}
\[
F(p , T - 2p -1) \sim\frac{(\mathit{SSR}_{\mathrm{restricted}} - \mathit{SSR}_{\mathrm{unrestricted}}) / p
}{\mathit{SSR}_{\mathrm{unrestricted}} / (T - 2p - 1)}.\vspace*{6pt}
\]
If this $F$ statistic is greater than the critical value for a chosen
level of
significance, we reject the null hypothesis that $X_t$ has no effect on
$Y_t$ and
conclude that $X_t$ Granger causes $Y_t$.
\subsection{Cointegration}
In Section \ref{trends}, we introduced the model of random walk with
drift as follows:
\begin{equation}
\label{random} Y_t = \beta_0 + Y_{t-1} +
u_t .
\end{equation}
If $Y_t$ follows Eq.\ \eqref{random}, then it has an autoregressive
root that equals 1.
If we consider a random walk for the first difference of the trend,
then we obtain
\begin{equation}
\label{difference} \Delta Y_t = \beta_0 + \Delta
Y_{t-1} + u_t .
\end{equation}
Hence, if $Y_t$ follows Eq.\eqref{difference}, then $\Delta Y_t$
follows a random walk, and accordingly \mbox{$\Delta Y_t - \Delta Y_{t-1}$} is
stationary; this is the second difference of $Y_t$ and is denoted
$\Delta^2 Y_t$. A series that has a random walk trend is said to be
integrated of order one, or I(1);\vadjust{\eject} a~series that has a trend of the form
\eqref{difference} is said to be integrated of order two, or I(2); and
a series that has no stochastic trend and is stationary is said to be
integrated of order zero, or I(0).
The order of integration in the I(1) and I(2) terminology is the number
of times that the series needs to be differenced for it to be
stationary. If $Y_t$ is I(2), then $\Delta Y_t$ is I(1), so $\Delta
Y_t$ has an autoregressive root that equals 1. If, however, $Y_t$ is
I(1), then $\Delta Y_t$ is stationary. Thus, the null hypothesis that
$Y_t$ is I(2) can be tested against the alternative hypothesis that
$Y_t$ is I(1) by testing whether $\Delta Y_t$ has a unit autoregressive
root. Sometimes, two or more series have the same stochastic trend in
common. In this special case, referred to as cointegration, regression
analysis can reveal long-run relationships among time series variables.
One could
think that a~linear combination of two processes I(1) is a process
I(1). However, this is not always true. Two or more series that have a
common stochastic trend are said to be \emph{cointegrated}. Suppose
that $X_t$ and $Y_t$ are integrated of order one. If, for some
coefficient $\theta$, $Y_t - \theta X_t$ is integrated of order zero,
then $X_t$ and $Y_t$ are said to be \emph{cointegrated}, and the
coefficient $\theta$ is called the \emph{cointegrating coefficient}. If
$X_t$ and $Y_t$ are cointegrated, then they have a common stochastic
trend that can be eliminated by computing the difference $Y_t - \theta
X_t$, which eliminates this common stochastic trend. There are three
ways to decide whether two variables can be plausibly modeled
exploiting the cointegration approach, namely, by expert knowledge and
economic theory, by a qualitative (graphical) analysis of the series
checking for common stochastic trend, and by performing statistical
tests for cointegration. In particular, there is a cointegration test
when $\theta$ is unknown. Initially, the cointegrating coefficient
$\theta$ is estimated by OLS estimation of the
regression\looseness=1
\begin{equation}
\label{coint} Y_t = \alpha+ \theta X_t +
z_t ,
\end{equation}
and then we use the Dickey--Fuller test (see Section \ref{trends})
to test for a unit root in $z_t$; this procedure is called the
Engle--Granger augmented Dickey--Fuller test for cointegration (EG-ADF
test); for details, see, for example, \cite[Chap.~6.2]{rif1} .
The concepts covered so far can be extended to the case of more than
two variables, for example, three variables, each of which is I(1), are
said to be cointegrated if $Y_t - \theta_1 X_{1t} - \theta_2 X_{2t}$ is
stationary.
The Dickey--Fuller needs the use of different critical values (see
Table~\ref{EG}), where the appropriate line depends on the number of
regressors used in the first step of estimating the OLS cointegrating
regression.

\begin{table}[!]
\caption{Critical values for the EG-ADF statistic}\label{EG}
\begin{tabular}{cccc}\hline
Numbers  of  regressors & $10\%$ & $5\%$ & $1\%$ \\\hline
1 & $-$3,12 & $-$3,41 & $-$3,96 \\
2 & $-$3,52 & $-$3,80 & $-$4,36 \\
3 & $-$3,84 & $-$4,16 & $-$4,73 \\
4 & $-$4,20 & $-$4,49 & $-$5,07 \\\hline
\end{tabular}
\end{table}

A different estimator of the cointegrating coefficient is the dynamic
OLS (DOLS) estimator, which is based on the equation
\begin{equation}
\label{DOLS} Y_t = \beta_0 + \theta X_t +
\sum_{j=-p}^p \delta_j
X_{t-j} + u_t .
\end{equation}
In particular, from Eq. \eqref{DOLS} we notice that DOLS includes past,
present, and future values of the changes in $X_t$. The DOLS estimator
of $\theta$ is the OLS estimator of $\theta$ in Eq.~\eqref{DOLS}. The
DOLS estimator is efficient, and statistical inferences about $\theta$
and $\delta$s in Eq.~\eqref{DOLS} are valid. If we have cointegration
in more than two variables, for example, three variable $Y_t, X_{1t},
X_{2t}$, each of which is I(1), then they are cointegrated with
cointegrating coefficients $\theta_1$ and $\theta_2$ if $Y_t - \theta_1
X_{1t} - \theta_2 X_{2t}$ is stationary. The EG-ADF procedure to test
for a single cointegrating relationship among multiple variables is the
same as for the case of two variables, except that the regression in
Eq.~\eqref{coint} is modified so that both $X_{1t}$ and $X_{2t}$ are
regressors. The DOLS estimator of a single cointegrating relationship
among multiple $X$s involves the level of each $X$ along with lags of
the first difference of each $X$.

\section{Conclusion}
In this first part of our ambitious project to use multivariate
statistical techniques to study critic econometric data of one of the
most influential economy in Italy, namely the Verona import--export
time series, we have focused ourselves on a self-contained introduction to
techniques of estimating OLS-type regressions, analysis of the
correlations obtained between the different variables and various types
of {\it information criteria} to check for the goodness of fit. A
particular relevance has been devoted to the application of tests able
to enlightening various types of nonstationarity for the considered
time series, for example, the {\it augmented Dickey--Fuller test} (ADF)
and the {\it Quandt likelihood ratio statistic} (QLR). Moreover, we
have also exploited both the {\it Granger causality} test and the {\it
Engle--Granger augmented Dickey--Fuller} test for cointegration
(EG-ADF) in order to analyze if and how these variables are related to
each other and to have a measure on how much a variable gives
information on the other one.
Such approaches constitute the core of the second part of our project,
namely the aforementioned Verona case study.

\section*{Acknowledgements}
The author would like to acknowledge the excellent support that Dr.
Chiara Segala gave him. Her help has been fundamental to develop the
whole project, particularly, for the realization of the applied
sections, which constitute the core of the whole work.


%
\end{document}